\documentclass[twocolumn,amsmath,amssymb,superscriptaddress]{revtex4-2}
\usepackage[dvipdfmx]{graphicx}
\usepackage{caption}
\usepackage{amsmath,amssymb}
\usepackage{amsthm}
\usepackage{ascmac}
\usepackage{subcaption}
\usepackage{float}
\usepackage{physics}
\usepackage{comment}
\usepackage[normalem]{ulem}

\usepackage{comment}

\usepackage{color}
\usepackage{hyperref}

\newcommand {\nn} {\nonumber}

\newcommand{\ee}{\mathrm{e}}
\newcommand{\ii}{\mathrm{i}}
\newcommand{\cut}{\mathrm{cut}}

\begin{document}
\begin{titlepage}

  \title{Quantum decoherence from complex saddle points}

\author{Jun N{\sc ishimura}}
 \email{jnishi@post.kek.jp}
\affiliation{KEK Theory Center,
Institute of Particle and Nuclear Studies,\\
High Energy Accelerator Research Organization,\\
1-1 Oho, Tsukuba, Ibaraki 305-0801, Japan}
\affiliation{Graduate Institute for Advanced Studies, SOKENDAI,\\
1-1 Oho, Tsukuba, Ibaraki 305-0801, Japan}

\author{Hiromasa W{\sc atanabe}}
\email{hiromasa.watanabe@yukawa.kyoto-u.ac.jp}
\affiliation{KEK Theory Center,
Institute of Particle and Nuclear Studies,\\
High Energy Accelerator Research Organization,\\
1-1 Oho, Tsukuba, Ibaraki 305-0801, Japan}
\affiliation{Yukawa Institute for Theoretical Physics, Kyoto University, Kyoto 606-8502, Japan}

\date{\today; preprint: KEK-TH-2648, YITP-24-104}

\begin{abstract}
\noindent  Quantum decoherence
  is the effect that bridges quantum physics
  to well-understood classical physics.
  As such,
  it plays a crucial role
  in understanding the mysterious nature of
  quantum physics.
  Quantum decoherence is also a source of quantum noise that has to be
  well under control in quantum computing and in
  various experiments based on quantum technologies.
  Here we point out that quantum decoherence can be
  captured by \emph{complex} saddle points in the Feynman path integral
  in much the
  same way as quantum tunneling can be captured by instantons.
  In particular,
  we present some first-principle calculations in the Caldeira--Leggett
  model, which
  reproduce the predicted scaling behavior
  of quantum decoherence with respect to the parameters of the environment
  such as the temperature and the coupling to the system of interest.
  We also discuss how to extend our approach to
  general models by Monte Carlo calculations using a recently developed method
  to overcome the sign problem.  
  \end{abstract}
\maketitle
\end{titlepage}

\textit{Introduction.---}
It is widely recognized that
quantum theory is a foundation of all the modern physics.
On the other hand,
there have been a lot of
confusions and debates about its mysterious nature.
One of the keys
to understand
this theory
is the quantum decoherence, which bridges
quantum physics to well-understood classical
physics.
(See \emph{e.g.}, Refs.~\cite{Schlosshauer,Zurek:1991vd}.)
More on the pragmatic side, one has to control
the quantum decoherence in order to develop a reliable quantum computer
and to perform experiments such as the gravitational wave detection,
which use quantum technologies.
For these reasons, it is important to be able to calculate
the effects of quantum decoherence explicitly
in a wide parameter region of various models.

A common strategy for studying a system coupled to some environment
is to use the master equation \cite{Lindblad:1975ef,Gorini:1975nb}
that describes the non-unitary time evolution
of the reduced density matrix of the system
after tracing out the environment.
(See also Section 4 of Ref.~\cite{Schlosshauer} and references therein.)
However,
master equations are obtained in general
only under some assumptions such as high temperature
together with some approximations
such as the Born and Markov approximations.
It is clearly desirable to develop alternative methods that do not rely
on such assumptions and approximations.

As more rigorous methods, one may think of investigating
the unitary time evolution of the whole system
including the environment.
For instance, one may attempt to solve the Schr\"odinger equation 
or to diagonalize the Hamiltonian.
(See, for instance, Refs.~\cite{Nagele:2020kef,adami_negulescu_2012}.)
However, the required computational cost grows exponentially
with the number of degrees of freedom in the whole system.

In this Letter, we investigate
the unitary time evolution of the whole system
by evaluating
the Feynman path integral explicitly.
In particular, we point out that
quantum decoherence can be captured by saddle points
in the real-time path integral formalism.
At first sight, this might look strange in light of the fact that the saddle point equation
derived from the action is nothing but the classical equation of motion,
whose real solution gives the classical motion.
In fact, by saddle points, we mean those including
the information of the initial quantum state, and hence they are complex in general.
[See Eq.\eqref{eq:saddle-pt_eq} and below.]
Thus in some sense, our finding
is analogous to the well-known fact that
quantum tunneling can be captured by instantons~\cite{coleman_1985},
which are \emph{real} saddle points in the \emph{imaginary}-time path integral.
See also Ref.~\cite{Nishimura:2023dky}
for a new picture of quantum tunneling
in terms of \emph{complex} saddle points
in the \emph{real}-time path integral.

Here we focus on
the Caldeira--Leggett (CL) model  \cite{Caldeira:1982iu,Caldeira:1982uj},
which has been studied intensively
as a model of quantum decoherence \cite{Paz:1992pn,Unruh:1989dd,Zurek:1992mv}.
(See 
Refs.~\cite{Zurek:2003zz,Schlosshauer:2019ewh} for reviews.)
The calculations simplify drastically in this case since
the path integral
to be evaluated
for typical initial conditions
is nothing but a multi-variable Gaussian integral.
  For instance, the reduced density matrix after some time evolution
  can be calculated \emph{exactly} by just obtaining the complex saddle points.
  This amounts to solving the saddle-point equation,
  which is a linear equation with
  a sparse complex-valued coefficient matrix.
  %
  %

  After identifying the parameter to be fixed in the limit of infinitely many
  degrees of freedom
  in the environment, we compare our results with the prediction from the master equation.
  While the use of master equation is not fully justified in the parameter region explored in this work,
  we observe qualitative agreement with the predicted scaling behavior of quantum decoherence
  with respect to the temperature and the coupling to the system of interest.
  The overall factor of the scaling behavior turns out to be larger
  than the predicted value by 40\%,  which clearly deserves further investigations.
  


\textit{Discretizing the Caldeira--Leggett model.---}
The Lagrangian used in our work is given by \cite{Caldeira:1982iu,Caldeira:1982uj}
\begin{align}
&    L  =  L_\mathcal{S} + L_\mathcal{E}  + L_\mathrm{int} \ ,
          \label{eq:Lagrangian}
          \\
&    L_\mathcal{S}
    =
    \frac{1}{2}\, M \dot{x}(t)^2
    - \frac{1}{2} \,  M \omega_\mathrm{b}^2 \, x(t)^2 \ ,
    \nn \\
&    L_\mathcal{E}
    =
    \sum_{k=1}^{N_\mathcal{E}}
    \left\{ \frac{1}{2} \, m \, \dot{q}^k(t)^2
    -\frac{1}{2} \, m \, \omega_k^{\;2} \, q^k(t)^2  \right\} \ ,
    \nn
      \\
&    L_\mathrm{int}
    =
    c\ x(t) \sum_{k=1}^{N_\mathcal{E}} \,q^k(t) \ ,
    \nn 
\end{align}
where we denote the coordinate of the $k$-th harmonic oscillator
as $q^k(t)$.
Let us note that
the mass parameters $M$ and $m$ in
\eqref{eq:Lagrangian}
can be absorbed
by rescaling $x \to x/ \sqrt{M}$, $q^k \to q^k/\sqrt{m}$ and
$c \to c \sqrt{Mm}$.
Hence, in what follows, we set $M=m=1$ without loss of generality.
%



The frequencies $\omega_k$ of the harmonic oscillators
in the environment are
determined
as follows.
Let us introduce a function
$\omega=g(\kappa)$
of $\kappa = \frac{k}{N_\mathcal{E}}$, which gives
$\dd \omega = (\dd g / \dd \kappa) \, \dd \kappa$.
Since the distribution of the harmonic oscillators
with respect to $\kappa$ is uniform,
the Ohmic spectrum \cite{Schlosshauer}
is reproduced if
\begin{align}
\left( \frac{\dd g}{\dd \kappa} \right)^{-1} \propto \omega^2 = g(\kappa)^2 \ ,
\end{align}
which implies $g(\kappa) \propto \kappa^{1/3}$.
Thus we obtain
\begin{equation}
    \omega_k  = 
    \omega_\mathrm{cut}
    \left(\frac{k}{N_\mathcal{E}}\right)^{1/3} \ ,
    \label{eq:omega_k_finiteNenv}
\end{equation}
where $\omega_\mathrm{cut}$ is the cutoff parameter.

In order to determine the coupling constant $c$ at finite $N_\mathcal{E}$,
let us complete the square
with respect to $q^k$ in the Lagrangian \eqref{eq:Lagrangian} as
\begin{equation}
    L  = \frac{1}{2} \, \dot{x}^2
    -
    \frac{1}{2}\, \omega_\mathrm{r}^2 x^2
    +
    \sum_{k=1}^{N_\mathcal{E}}
    \qty[
    \frac{1}{2} \,  (\dot{q}^k)^2
    -
    \frac{1}{2} \, \omega_k^2 \qty(q^k - \frac{c}{\omega_k^2}x)^2
    ] \ ,
    \label{eq:Lagrangian_coupled_osc}
\end{equation}
where we have defined the renormalized frequency $\omega_\mathrm{r}$ by
\begin{equation}
    \omega_\mathrm{r}^2 
    =  \omega_\mathrm{b}^2 -  c^2
    \sum_{k=1}^{N_\mathcal{E}}  \frac{1}{\omega_k^2} \ ,
    \label{eq:omega_ren_discrete}
\end{equation}
as opposed to the bare frequency $\omega_\mathrm{b}$.
Since the harmonic oscillators $q^k$ in the environment
are expected to oscillate around the potential minimum
$cx/\omega_k^2$ when $x$ varies slowly with time,
the frequency $\omega_\mathrm{b}$
of the system $\mathcal{S}$ is shifted to
\eqref{eq:omega_ren_discrete}
due to the environment $\mathcal{E}$.
We identify \eqref{eq:omega_ren_discrete} with the formula
\begin{align}
  \label{eq:omega_ren_large_N}
  \omega_\mathrm{r}^2
  =  \omega_\mathrm{b}^2
  - \frac{4\gamma \, \omega_\cut}{\pi}
\end{align}
derived
in the large $N_\mathcal{E}$ limit
(See \emph{e.g.}, Ref.~\cite{Schlosshauer}),
where $\gamma$ represents the effective coupling that appears
in the CL master equation.
Thus we obtain the relationship between $\gamma$ and the coupling constant $c$ as
\begin{equation}
    c^2     = 
    \frac{4\gamma}{\pi} \omega_\cut^3 
    \left\{
    \sum_{k=1}^{N_\mathcal{E}} \qty(\frac{N_\mathcal{E}}{k})^{2/3}
    \right\}^{-1} \ .
    \label{eq:coupling_scaling}
\end{equation}


%
%
%


In order to put the whole system on a computer,
we discretize the time $t$ as
$t_n = n \, \epsilon$ $(n=0,\cdots, N_t)$,
where $t_\mathrm{F} \equiv t_{N_t}$.
Accordingly, the variables
$x(t)$ and $q^k(t)$ are also discretized
as $x_n = x(t_n)$ and $q^k_n =q^k(t_n)$.
The action with the discretized time can be written as
\begin{align}
&  S(x,q)
    = 
    \frac{1}{2} \, \epsilon
    \sum_{n=0}^{N_t-1}
    \qty[
    \qty(\frac{x_n-x_{n+1}}{\epsilon})^2
    -
    \omega_\mathrm{b}^2
    \, 
    \frac{x_n^2+x_{n+1}^2}{2}
    ]
\nn      \\
& +  \frac{1}{2} \, \epsilon
    \sum_{k=1}^{N_\mathcal{E}}
    \sum_{n=0}^{N_t-1}
    \qty[
    \qty(\frac{q_n^k-q_{n+1}^k}{\epsilon})^2
    -
    \omega_k^2
    \, 
    \frac{(q_n^k)^2+(q_{n+1}^k)^2}{2}
    ]
\nn     \\
    & +
c \, \epsilon 
 \sum_{k=1}^{N_\mathcal{E}} \sum_{n=0}^{N_t-1}
\frac{x_n q_n^k+x_{n+1} q_{n+1}^k}{2}
\ .
    \end{align}
%
%
%


We assume that the initial condition for the density matrix is given by
\begin{equation}
    \hat{\rho}(t=0) 
    =
    \hat{\rho}_\mathcal{S}(t=0)\otimes\hat{\rho}_\mathcal{E}  \ .
\end{equation}
As the initial density matrix $\hat{\rho}_\mathcal{S}(t=0)$ of the system $\mathcal{S}$,
we consider a pure state with the Gaussian wave packet
\begin{align}
    \rho_\mathcal{S}(x,\tilde{x};t=0) 
    &=
    \psi_\mathrm{I}
    (x)\psi^* _\mathrm{I}
    (\tilde{x}) \ ,
    \label{init-rho-system-gen}
    \\
    \psi
    _\mathrm{I}
    (x)  &= 
    \exp(-\frac{1}{4 \sigma^2} x^2) \ .
    \label{init-rho-system-gaussian}
\end{align}

\begin{figure}[t]
	\centering
	\includegraphics[width=0.4\textwidth]{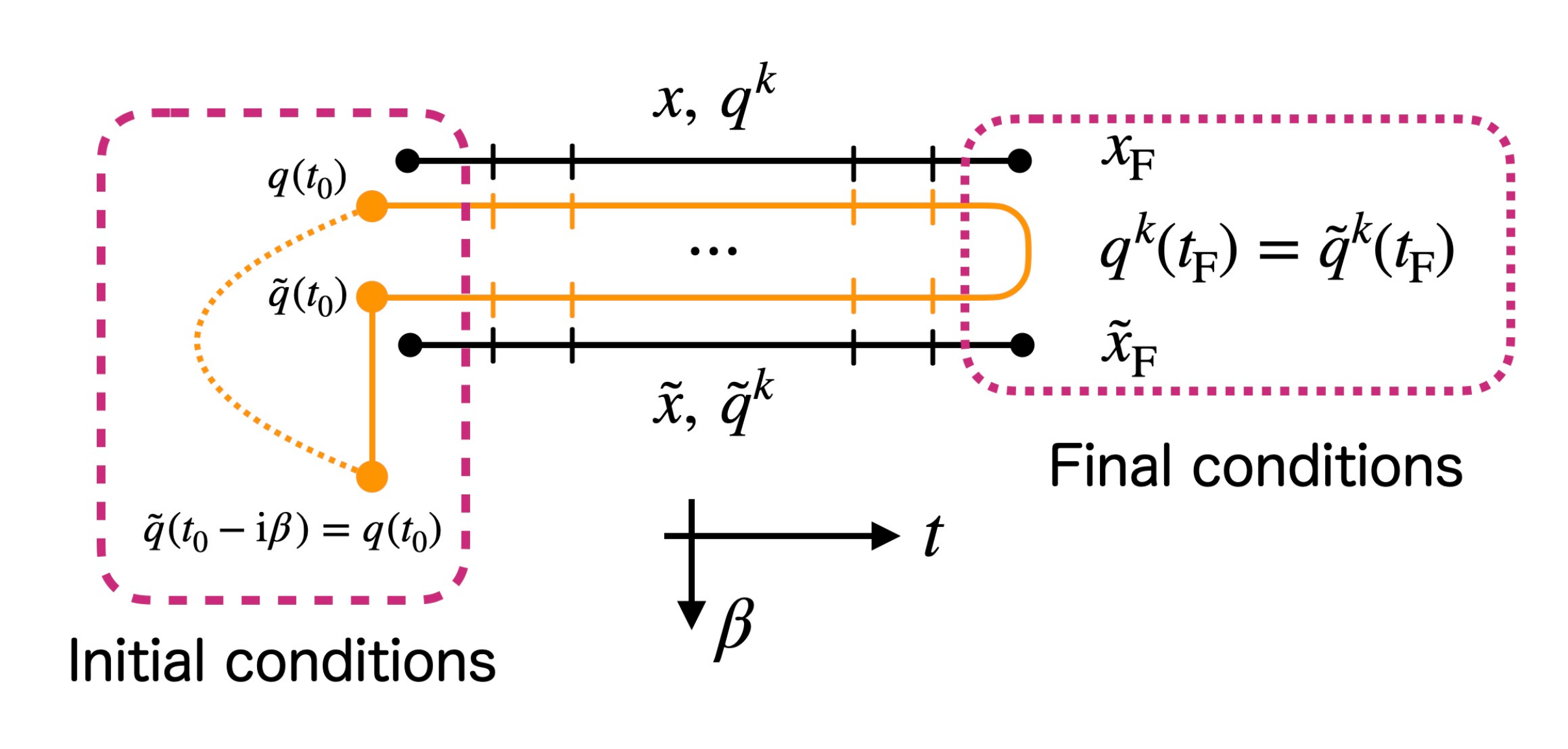}
	\caption{
          Schematic picture of the path integral \eqref{eq:rho_S_position_basis-summary}
          used to calculate the reduced density matrix
$\rho_\mathcal{S}(x_\mathrm{F},\tilde{x}_\mathrm{F};t_\mathrm{F})$
          of the system.
          The boundary conditions are imposed
          at the initial time and the final time.
	}
        \label{fig:lattice-reduced-density-matrix}
\end{figure}

As the initial density matrix $\hat{\rho}_\mathcal{E}$ of the environment $\mathcal{E}$,
we take
the canonical ensemble with the temperature $T \equiv \beta^{-1}$.
For that,
we introduce an additional path
for the variables $\tilde{q}^k$ in
the imaginary time direction
with
the free Euclidean action
(See Fig.~\ref{fig:lattice-reduced-density-matrix})
\begin{align}
&    S_0(\tilde{q}) = 
\frac{1}{2} \, \tilde{\epsilon}
    \sum_{k=1}^{N_\mathcal{E}}
    \sum_{j=0}^{N_\beta-1}
    \left[
        \qty(\frac{\tilde{q}^k_0(j+1)-\tilde{q}^k_0(j)}{\tilde{\epsilon}})^2 \right.
\nn        \\
& \quad \quad \quad \quad \quad \quad \quad \quad \quad 
 + \,   \omega_k^2
 \, 
 \frac{\tilde{q}^k_0(j+1)^2 + \tilde{q}^k_0(j)^2}{2}
 \Big] \ ,
\end{align}
where we define $\tilde{q}^k_0(j) = \tilde{q}^k(t_0-\ii(j\tilde{\epsilon}))$
and impose $q^k(t_0) =\tilde{q}^k(t_0 - \ii \beta)$
with
$\beta = N_\beta \, \tilde{\epsilon}$,
namely $q^k_0= \tilde{q}^k_0(N_\beta)$.


Thus the reduced density matrix
of the system $\mathcal{S}$
can be given by
\begin{equation}
    \rho_\mathcal{S}(x_\mathrm{F},\tilde{x}_\mathrm{F};t_\mathrm{F})
    = \!\! \int \!\! \mathcal{D}x\mathcal{D}\tilde{x}
    \prod_{k=1}^{N_\mathcal{E}} \mathcal{D}q^k\mathcal{D}\tilde{q}^k \mathcal{D}\tilde{q}_0^k
    \, \mathrm{e}^{-S_{\rm eff}(x,\tilde{x},q,\tilde{q},\tilde{q}_0)}
    \ ,
    \label{eq:rho_S_position_basis-summary}
\end{equation}
whose elements are
specified by the boundary condition
for the system $\mathcal{S}$ as
$x(t_\mathrm{F}) = x_\mathrm{F}$ and $\tilde{x}(t_\mathrm{F}) = \tilde{x}_\mathrm{F}$
at the final time.
Corresponding to taking the trace with respect to the environment $\mathcal{E}$,
we also impose $q^k(t_\mathrm{F}) =\tilde{q}^k(t_\mathrm{F})$.
The effective action
in \eqref{eq:rho_S_position_basis-summary}
is
given by
\begin{align}
  S_{\rm eff} (x,\tilde{x},q,\tilde{q},\tilde{q}_0)
  &= - \mathrm{i} \left\{ S(x, q) - S(\tilde{x}, \tilde{q}) \right\}
  \nn \\
  & \quad
  + S_0(\tilde{q}_0) + \frac{1}{4 \sigma^2} (x_0 ^2 + \tilde{x}_0 ^2)  \ .
  \label{def-eff-action}
\end{align}
Note that the first two terms, which are purely imaginary, represent the original action, and
the last two terms, which are real, represent the initial quantum state.
Since the integrand of \eqref{eq:rho_S_position_basis-summary}
is complex, the path integral becomes highly oscillatory, which
makes ordinary Monte Carlo calculations inapplicable due to severe cancellations among
generated configurations. This is well known as the sign problem.
(See Ref.~\cite{Alexandru:2015sua}, for instance.)


\textit{Performing the path integral.---}
In the present case, the effective action \eqref{def-eff-action} is
quadratic with respect to
the integration variables, and it can be written as
\begin{align}
  S_{\rm eff} (x,\tilde{x},q,\tilde{q},\tilde{q}_0)
    = 
    \frac{1}{2}&X_\mu \mathcal{M}_{\mu\nu}X_\nu - C_\mu X_\mu  + B \ ,
    \label{eq:primitive_lattice_action}
\end{align}
where $X_\mu$ ($\mu = 1 , \cdots , D$) represents the integration variables collectively
and the number of integration variables is
$D = 2N_t (1 + N_\mathcal{E})  + N_\beta N_\mathcal{E}$.
Note that $\mathcal{M}$ is a $D\times D$  complex symmetric matrix,
which is independent of $x_\mathrm{F}$ and $\tilde{x}_\mathrm{F}$, whereas
$C_\mu$ and $B$ are purely imaginary quantities defined by
\begin{align}
& C_\mu X_\mu  = 
   -  \frac{\mathrm{i}}{\epsilon}
   \qty(x_\mathrm{F}x_{N_t-1} - \tilde{x}_\mathrm{F}\tilde{x}_{N_t-1})
   \nn \\
   & \quad\quad \quad\quad  +   \frac{\mathrm{i}}{2} c \, \epsilon \sum_k
    \qty(x_\mathrm{F} - \tilde{x}_\mathrm{F}) q^k_{N_t} \ , 
    \notag \\
&   B =   -  \frac{\mathrm{i}}{2} \, b  \, (x_\mathrm{F}^2 - \tilde{x}_\mathrm{F}^2 )  \ ,
    \quad
    \mbox{~where~}
    b = \frac{1}{\epsilon}-\frac{\omega_\mathrm{b}^2\epsilon}{2}   \ .
    \label{eq:def-C-B}
\end{align}
Since $C_\mu$ is linear in $x_\mathrm{F}$ and $\tilde{x}_\mathrm{F}$,
let us write them as
\begin{align}
  C_\mu  &=  \mathrm{i} (c_\mu x_\mathrm{F} - \tilde{c}_\mu \tilde{x}_\mathrm{F} ) \ .
    \label{eq:def-c-mu}
\end{align}


The saddle point of this action
is given by
\begin{equation}
  \bar{X}_\mu = \left( \mathcal{M}^{-1} \right)_{\mu\nu}
  \, C_\nu \ ,
    \label{eq:saddle-pt_eq}
\end{equation}
which is complex in general, reflecting the fact that
the action \eqref{def-eff-action} is complex.
Note also that the saddle point includes the information of the
initial quantum state represented by the last two terms in \eqref{def-eff-action}.
For these reasons, the saddle point we obtain here should not be regarded
as something that represents the classical motion, which corresponds to 
a real saddle point derived \emph{solely} from the original action with some boundary conditions.
Rather, it is analogous to the complex saddle points
representing quantum tunneling in the real-time path integral \cite{Nishimura:2023dky},
which corresponds to the instantons \cite{coleman_1985} in the imaginary-time path integral
through analytic continuation.

Redefining the integration variables as $Y_\mu = X_\mu - \bar{X}_\mu$,
the effective action becomes
\begin{equation}
  S_{\rm eff} (x,\tilde{x},q,\tilde{q},\tilde{q}_0)
  =   \frac{1}{2}Y_\mu \mathcal{M}_{\mu\nu}Y_\nu
  + \mathcal{A} \ ,
\end{equation}
where we have defined
\begin{align}
\mathcal{A}
  &=  B - \frac{1}{2} \,  C_\mu  \left( \mathcal{M}^{-1} \right)_{\mu\nu}
  C_\nu
 \ .
\end{align}
Integrating out $Y_\mu$, we obtain
\begin{align}
  \rho_\mathcal{S}(x_\mathrm{F}, \tilde{x}_\mathrm{F};t_\mathrm{F})
  &= \frac{1}{\sqrt{\det \mathcal{M}}} \ee^{-\mathcal{A}} \ .
\label{eq:rho-detM}
\end{align}
%
%

Let us consider the magnitude
$| \rho_\mathcal{S}(x_\mathrm{F}, \tilde{x}_\mathrm{F};t_\mathrm{F}) |$,
which is determined by
\begin{align}
 {\rm Re } \mathcal{A}
  &=  \frac{1}{2}
  \begin{pmatrix}
x_\mathrm{F}  & \tilde{x}_\mathrm{F} 
\end{pmatrix}
  \begin{pmatrix}
    J   &   -K  \\
  -K   &  J   \\
\end{pmatrix}
\begin{pmatrix}
x_\mathrm{F}  \\ \tilde{x}_\mathrm{F} 
\end{pmatrix}
\\
&=  \frac{1}{4} \,
\{ (J-K)( x_\mathrm{F}  +  \tilde{x}_\mathrm{F}  )^2
+ (J+K)( x_\mathrm{F}  -  \tilde{x}_\mathrm{F}  )^2  \} \ ,
\end{align}
where we have defined
\begin{align}
  J =  {\rm Re} \{   c_\mu (\mathcal{M}^{-1})_{\mu\nu} c_{\nu} \} 
= {\rm Re} \{ \tilde{c}_\mu (\mathcal{M}^{-1})_{\mu\nu} \tilde{c}_{\nu}  \}  \  , \\
K = {\rm Re} \{ c_\mu (\mathcal{M}^{-1})_{\mu\nu} \tilde{c}_{\nu}\} =
  {\rm Re} \{ \tilde{c}_\mu (\mathcal{M}^{-1})_{\mu\nu} c_{\nu} \} \ .
    \label{eq:def-J-K}
\end{align}
Thus we obtain
\begin{align}
   | \rho_\mathcal{S}(x_\mathrm{F}, \tilde{x}_\mathrm{F};t_\mathrm{F}) |
& \simeq
  \exp \left\{ - \frac{1}{2}
  \Gamma_\mathrm{diag}(t_\mathrm{F})
  \left( \frac{x_\mathrm{F}  +  \tilde{x}_\mathrm{F}}{2}  \right)^2 \right.
  \nn \\
  & \quad \left.
  -  \frac{1}{2}    \Gamma_\mathrm{off\mathchar`-diag}(t_\mathrm{F})
 \left( \frac{x_\mathrm{F}  -  \tilde{x}_\mathrm{F}}{2}  \right)^2
 \right\} \ ,
    \label{eq:diff_logrho-def}
\end{align}
omitting the prefactor
independent of $x_\mathrm{F}$ and $\tilde{x}_\mathrm{F}$,
where we have defined the quantities
\begin{align}
  \Gamma_\mathrm{diag}(t_\mathrm{F}) &= 2 (J-K) \ , \\
  \Gamma_\mathrm{off\mathchar`-diag}(t_\mathrm{F}) &= 2 (J+K) \ ,
    \label{eq:def-Gamma}
\end{align}
which characterize the fall-off of the
matrix element
in the diagonal and off-diagonal directions, respectively.

A characteristic behavior of quantum decoherence is the disappearance 
of the off-diagonal elements of the reduced density matrix at early times,
which can be probed by the increase of
$\Gamma_\mathrm{off\mathchar`-diag}(t)$ with $t$.
In particular, the CL master equation predicts \cite{Schlosshauer,PhysRevD.45.2843} a linear growth
\begin{equation}
    \Gamma_\mathrm{off\mathchar`-diag}(t) = \frac{8\gamma}{\beta} \, t \ ,
    \label{eq:decoherence_from_Gamma_offdiag}
\end{equation}
at small $\gamma$ and small $\beta$ (high temperature).

\textit{Numerical results.---}
We consider the case in which
the initial state of
the system $\mathcal{S}$ is
chosen to be
the ground state of the harmonic oscillator with the renormalized frequency
$\omega_\mathrm{r}$, which corresponds
to setting $\sigma^2 = \frac{1}{2 \omega_\mathrm{r}}$
in
\eqref{init-rho-system-gaussian}.
At $t=0$,
the width of the Gaussian distribution is
$\Gamma_\mathrm{diag}(0) =
\Gamma_\mathrm{off\mathchar`-diag}(0)=2\, \omega_\mathrm{r}$.

Here we set 
$\omega_\mathrm{r} =0.08$,
$\omega_\cut= 2.0$, which satisfies
$\omega_\mathrm{r} \ll \omega_\cut$, and
use $N_\mathcal{E}=64$ \footnote{The bare frequency $\omega_\mathrm{b}$
is determined
by \eqref{eq:omega_ren_large_N},
whereas the coupling constant $c$ is determined by \eqref{eq:coupling_scaling}.}.
The lattice spacing in the time direction
is chosen to be $\epsilon=0.05$, whereas 
the lattice spacing in the temperature direction
is chosen to be
$\tilde{\epsilon}=0.05$ for $\beta\ge 0.2$,
and $\tilde{\epsilon}=\beta/4$ for $\beta\le 0.2$.


\begin{figure}[t]
  \centering
\includegraphics[width=0.45\textwidth]{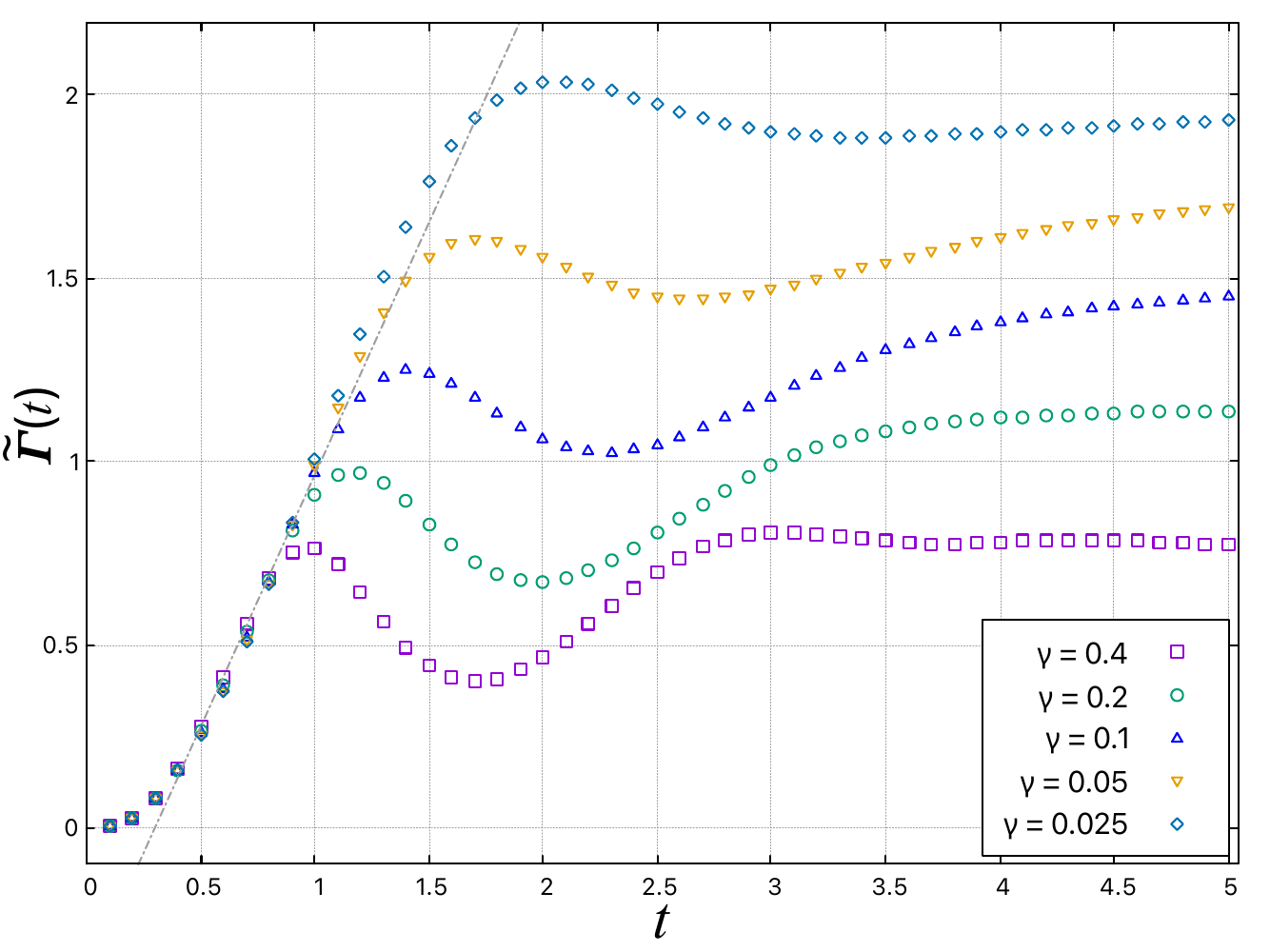}
\caption{The rescaled quantity \eqref{eq:rescaled-Gamma}
          is plotted against $t$
          for $\gamma=0.025$, $0.05, \cdots , 0.4$ with
          $\beta=0.05$.
          The dash-dotted line represents a fit of the $\gamma=0.1$ data
          within $0.4 \le t \le 1.1$
          to a linear behavior $At+B$, where $A\sim 1.38$ is obtained.}
        \label{fig:gamma_dependency}
\end{figure}

\begin{figure}[t]
  \centering
%
    \includegraphics[width=0.45\textwidth]{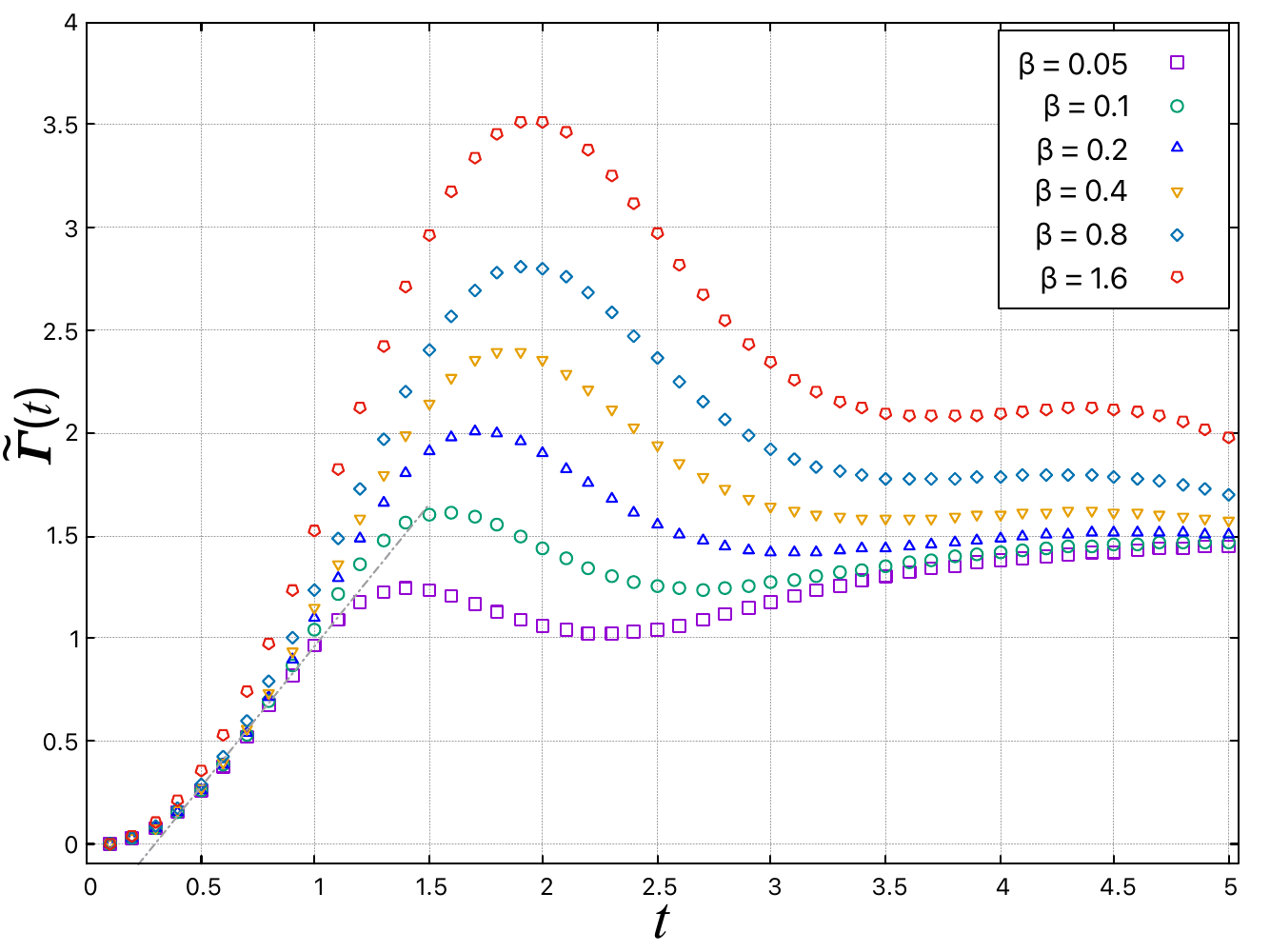}
    \caption{The
rescaled quantity \eqref{eq:rescaled-Gamma}
    is plotted against $t$
    for $\beta = 0.05$, $0.1$, $0.2 , \cdots , 1.6$
    with
    $\gamma = 0.1$.
    The dash-dotted line represents a fit of the $\beta=0.05$ data
              within $0.4 \le t \le 1.1$
    to a linear behavior $At+B$, where $A\sim 1.38$ is obtained.}
        \label{fig:T_dependency}
\end{figure}

In Fig.~\ref{fig:gamma_dependency},
we plot the rescaled quantity
\begin{align}
  \tilde{\Gamma}(t) &= \frac{\beta}{8\gamma} \, \{\Gamma_\mathrm{off\mathchar`-diag}(t)
    -\Gamma_\mathrm{off\mathchar`-diag}(0)\}
    \label{eq:rescaled-Gamma}
\end{align}
against $t$
for $\gamma=0.025$, $0.05, \cdots , 0.4$ with
$\beta=0.05$,
    which reveals a nice scaling behavior at early times.
The scaling behavior implies
$\Gamma_\mathrm{off\mathchar`-diag}(t) = \frac{8\gamma}{\beta} (A t + B)$,
which is qualitatively consistent with the prediction from the master equation.
However, the overall factor
$A\sim 1.38$ obtained by the fit to the $\gamma=0.1$ data
is slightly larger than the predicted value $A=1$.
In fact, we find that the fitted value of $A$ depends on the choice of 
$\omega_\mathrm{cut}$, which suggests that 
the separation $\omega_\mathrm{cut} \ll T=\beta^{-1}$
may not be good enough to justify the prediction based on the master
equation. (See Fig.~5 of Ref.~\cite{Paz:1992pn} for analogous results.)
We therefore consider that precise agreement should be obtained
by taking the 1) $T\rightarrow \infty$, 2) $N_\mathcal{E}\rightarrow \infty$
and 3) $\omega_\mathrm{cut} \rightarrow \infty$ limits carefully.

In Fig.~\ref{fig:T_dependency},
we plot
the rescaled quantity \eqref{eq:rescaled-Gamma}
against $t$
for $\beta = 0.05$, $0.1 , \cdots , 1.6$
with
$\gamma = 0.1$,
which reveals a nice scaling behavior at early times.
However, the linear behavior is not clearly seen except for $\beta=0.05$.
It is conceivable that
the high temperature limit assumed in the prediction
based on the master equation is not valid for $\beta \gtrsim 0.1$.

Details of our calculation are reported in 
a separate paper \cite{Nishimura:2025dam}.
There we also perform calculations
for various $N_\mathcal{E}=8, 16, \cdots , 256$
with $\beta = 0.05$ and $\gamma = 0.1$.
We see a clear converging behavior to $N_\mathcal{E}=\infty$
for $t\lesssim 3$,
which confirms the validity of our choice \eqref{eq:coupling_scaling}
of the coupling constant $c$ for finite $N_\mathcal{E}$.




\textit{Discussions.---}
In this Letter we have pointed out that quantum decoherence
can be captured by complex saddle points in the
real-time path integral formalism,
which may be taken as a surprise
given that quantum decoherence is
a genuinely quantum effect,
which is expected to provide a key to link quantum theory to classical theory.
We consider that our finding is important from both theoretical and practical
points of view.
On the theoretical side, the complex saddle-point
configurations may provide a conceptual
understanding
of quantum decoherence just like instantons do for quantum tunneling.
On the practical side, our finding suggests a whole new approach to quantum decoherence
based on the saddle point analysis or its extension using Monte Carlo methods
as we discuss below.

In order to substantiate our assertion,
we have investigated the CL model with typical initial conditions,
where we are able to obtain exact results for arbitrary values of the
parameters such as the number of harmonic oscillators $N_\mathcal{E}$,
the coupling constant $\gamma$ and the temperature $T=\beta^{-1}$.
In particular, we have compared our results
with those obtained by the master equation,
and observe qualitative agreement in the scaling behavior with respect to
$\gamma$ and $\beta$.
Let us emphasize, however, that our approach does not use any
assumptions or approximations, and hence it can be applied to
arbitrary coupling constant and temperature.

While we have focused on the initial decoherence rate in this Letter,
in the separate paper \cite{Nishimura:2025dam},
we have also examined the late-time asymptotics
of the widths
$\Gamma_\mathrm{diag}(t)$ and $\Gamma_\mathrm{off\mathchar`-diag}(t) $
of the reduced density matrix
towards the equilibrium as we increase $N_\mathcal{E}$ of the thermal environment
from 8 to 256.
This can be contrasted to the common approach,
where the $N_\mathcal{E}\to\infty$ limit has to be taken in order to derive
the master equation.
Thus our method serves as a complementary tool in investigating issues
related to thermalization as well.
%
In the same paper \cite{Nishimura:2025dam},
we have also generalized our approach to
the initial state with
a superposition of two Gaussian wave packets.
Quantum decoherence in that case can be seen more dramatically
as the fading of the interference pattern.

In fact, one can obtain exact results from saddle points for any model with
a Gaussian action and a Gaussian initial state in the way we have done in this work.
Despite the simplicity of the setup,
one can actually investigate
various behaviors of quantum many body systems other than
quantum decoherence such as dissipation and
thermalization.

In more general models, one needs to perform the real-time path integral
that goes beyond the Gaussian integral. In fact,
even finding all the complex saddle points that contribute to the
path integral is not straightforward.
(Such solutions are referred to as \emph{relevant} saddle points in the literature.)
Here we propose to use the recently developed Monte Carlo method called
the generalized Lefschetz thimble method (GTM) \cite{Alexandru:2015sua}.
This method enables us not only to identify all the relevant complex saddle points
but also to perform numerical integration around each saddle point 
along the so-called Lefschetz thimble,
which is nothing but a multi-dimensional version of the
steepest descent path in the saddle point analysis.

More precisely,
the GTM is a method that has been proposed to overcome the sign problem
that occurs in Monte Carlo calculations when the integrand of the path integral
is not positive semi-definite.
The idea is to
complexify the integration variables
and to deform the integration contour based on Cauchy's theorem
in such a way that the sign problem is
ameliorated.
(See Refs.~\cite{Witten:2010cx,Cristoforetti:2012su,Cristoforetti:2013wha,Fujii:2013sra}
for earlier proposals to perform integration precisely on the Lefschetz thimbles.).
In particular,
various important techniques 
developed more recently \cite{Fukuma:2017fjq,Fukuma:2019uot,Fukuma:2020fez,Fukuma:2021aoo,Fujisawa:2021hxh,Nishimura:2024bou},
enabled, for instance,
the investigation of quantum tunneling \cite{Nishimura:2023dky}
and quantum cosmology \cite{Chou:2024sgk}
based on the real-time path integral,
where the relevant saddle points and the associated Lefschetz thimbles
that contribute to the path integral
have been clearly identified.
Similarly, it is expected that this method is useful in investigating
a system coupled to the environment.
Our observation that quantum decoherence can be captured by
complex saddle points suggests that the GTM is particularly suitable
for investigating
such a system
from first principles.


\textit{Acknowledgements.---} We would like to thank Yuhma Asano,
Masafumi Fukuma, Kouichi Hagino, Yoshimasa Hidaka, Katsuta Sakai,
Hidehiko Shimada, Kengo Shimada, Hideo Suganuma and Yuya Tanizaki for
valuable discussions and comments. H.~W.\ was partly supported by Japan
Society for the Promotion of Science (JSPS) KAKENHI Grant numbers,
21J13014 and 23K22489.



\bibliography{ref}


\end{document}